\documentclass[a4paper, 12pt]{article}
\usepackage{setspace}
\usepackage{latexsym}
\usepackage{graphicx}
\usepackage{epsfig}

\begin{document}
\newcommand{\eg}{{\it e.g.}}
\newcommand{\etal}{{\it et. al.}}
\newcommand{\ie}{{\it i.e.}}
\newcommand{\be}{\begin{equation}}
\newcommand{\dd}{\displaystyle}
\newcommand{\ee}{\end{equation}}
\newcommand{\bea}{\begin{eqnarray}}
\newcommand{\eea}{\end{eqnarray}}
\newcommand{\bef}{\begin{figure}}
\newcommand{\eef}{\end{figure}}
\newcommand{\bce}{\begin{center}}
\newcommand{\ece}{\end{center}}
\newcommand{\bi}{\bibitem}
\newcommand{\bit}{\begin{itemize}}
\newcommand{\eit}{\end{itemize}}
\newcommand{\Z}{\mathbf{Z}}
\newcommand{\R}{\mathbf{R}}
\newcommand{\C}{\mathbf{C}}
\newcommand{\1}{\mathbf{1}}
\newcommand{\eps}{\epsilon}
\newcommand{\beps}{\bar\epsilon}
\newcommand{\bpsi}{\bar\psi}
\newcommand{\blambda}{\bar\lambda}
\newcommand{\bsigma}{\bar\sigma}
\newcommand{\dalpha}{\dot\alpha}
\newcommand{\dbeta}{\dot\beta}
\newcommand{\dgamma}{\dot\gamma}
\newcommand{\ddelta}{\dot\delta}
\newcommand{\lra}{\leftrightarrow}
\newcommand{\la}{\leftarrow}
\newcommand{\ra}{\rightarrow}
\newcommand \dsl {\not\!\partial}
\newcommand \dslbar {\not\!\bar\partial}

\def\lsim{\mathrel{\rlap{\lower4pt\hbox{\hskip1pt$\sim$}}
    \raise1pt\hbox{$<$}}}         
\def\gsim{\mathrel{\rlap{\lower4pt\hbox{\hskip1pt$\sim$}}
    \raise1pt\hbox{$>$}}}         

\thispagestyle{empty}
\begin{center}


%
%
%
%
%
%
%
%
%
%
%
%
%
%
%
%
%
%
%
%
%
%
%
%
%
%

{\Large \sc Charles University in Prague}

\vskip .5cm

{\Large \sc Faculty of Mathematics and Physics}

\vskip 2.5cm

{\huge \sc  Alternative Symmetries in}

\vskip .5cm

{\huge \sc Quantum Field Theory and}

\vskip .5cm

{\huge \sc Gravity}

\vskip 2cm

{\huge  Alfredo Iorio, PhD}

\vskip 5.5cm

{\huge \sc Habilitation Thesis for }

\vskip .5cm

{\huge \sc Associate Professor}

\vfill

{\large \sc May 2010}

\end{center}

\newpage

.

\newpage

\section{\sc Prologue} Let me begin by explaining the title. By ``alternative symmetries'' it is meant symmetries that in various ways are \textit{modifications} of established symmetries: Poincar\'e and gauge symmetries of flat space-times, diffeomorphisms of general relativity (GR), etc. The modifications I shall be dealing with here are either an \textit{enhancement}, like conformal symmetry and Supersymmetry (SUSY), or an \textit{impoverishment}, like the relic symmetries of a field theory built on a noncommutative space, or else a \textit{deformation}, like quantum groups (q-groups) that reduce to standard symmetry groups in a limit. Sometimes the modification comes about due to a \textit{dimensional reduction} from 3+1 dimensions to 2+1 and even 1+1 dimensions. This is the case of the appearance in 2+1 dimensions of a Chern-Simons (CS) gravitational term (or conformal gravity term) besides the usual Einstein-Hilbert (EH) term of GR.

\begin{figure}
 \centering
  \includegraphics[height=.4\textheight]{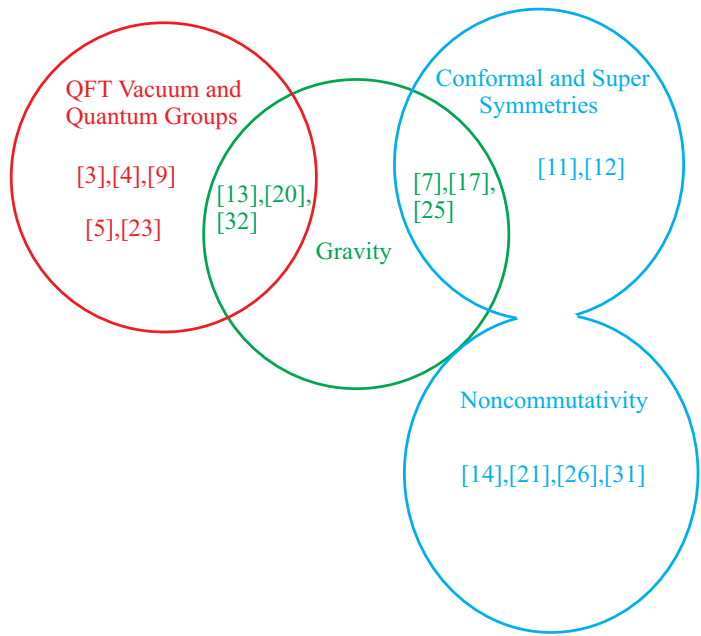}
  \caption{Schematic view of the discussed research areas. The references $[n]$ are to the 17 papers selected for inclusion in this Dissertation, see References. An effort is made to arrange the papers into homogeneous sets.}
\label{scheme}
\end{figure}

Following a path that is the intention of this Dissertation to describe, various aspects of the above - q-groups in relation to the vacuum structure of quantum field theory (QFT), conformal symmetry in flat as well as curved spaces, SUSY, noncommutative field theories (NCFTs), gravity and gravity-analog models - have been the focus of most of my investigation. This can be easily seen combing through the titles of the list of works in\footnote{In what follows the references to papers included in the last Section of this Dissertation are in bold typeface. See also Fig.~\ref{scheme}.} \cite{thesis1} -\cite{colorhorizon}. That list comprises two theses - the Laurea thesis \cite{thesis1} and the PhD thesis \cite{thesis2} - all papers, written-up talks  and other contributions to my understanding of various issues, including three sets of (incomplete) lecture notes \cite{lectnotes1}-\cite{lectnotes3}. The most recent work cited there is either just published \cite{dnaletter}, \cite{dnapaper} or in progress  \cite{dnatorus}-\cite{colorhorizon} and should be taken as a commitment to future research.

Although it seems self-explanatory, the rest of the title also deserves some elucidations. That is ``quantum field theory'' and
``gravity''. QFT is not only meant here in the traditional sense of the relativistic theory of quantum processes, but also in the more general sense of the theory of quantum systems with an infinite number of degrees of freedom - which means that the emphasis is on the field aspects - hence relativity might well be the Galilean one or something else\footnote{E.g., as some of my work is dedicated to NCFTs that break rotations and boosts invariance in a prescribed manner, the spatiotemporal symmetry group there is $SO(2) \times SO(1,1)$.}. This opens-up applications to areas that lay outside particle physics - the area where QFT was born - such as condensed matter physics and beyond, see, e.g., {\bf \cite{qQD}}, {\bf \cite{pla}}, {\bf \cite{shor}}. As for gravity, I would like to say that I work on non-canonical aspects of gravity, such as the relevance of the unitarily inequivalent representations (UIRs) of QFT for black-hole physics {\bf \cite{TFDBH}}, topologically non-trivial solutions of a sector of 2+1 dimensional gravity (the above mentioned conformal or CS gravity term) {\bf \cite{cs3}}, {\bf \cite{sucs}}, as well as applications of ideas born in a gravitational context to problems of different physical origin, such as hadronization in heavy-ion collisions {\bf \cite{rindler}} and such as topological defects in elastic media \cite{gaugevirus}.

This research activity has its habitat in what we might call an era of the search for physics beyond QFT and GR. This search started with QFT itself, the prominent direction probably being the search for a meaningful quantum theory of gravity. Decades later it still has to come to an end or, in some cases, even to a solid beginning. The theoretical mind-set established by this titanic struggle is to empower the experimentally tested symmetry principles to the extent of gaining a high level of mathematical control on the models but paying the price of introducing more and more exotic speculations (i.e. not testable by experiments) regarded as real.

The aware reader might have guessed that the models I am alluding to here are those born within string theory (ST). ST would safely fit the title of this Dissertation and, in a way, through SUSY or NCFTs, part of my research has tight links with it (see Section \ref{stsymm}). Nonetheless, in my work I attempted to proceed, on the one hand, by sharing the conviction that new and powerful concepts relevant for physics are hidden in these alternative views (see the first paragraph here), on the other hand, by holding to certain epistemological pillars usually absent from ST: to have space-time dimensions never greater than 4 (may be less, but not more); to propose experimental tests even of the most speculative ideas (such as noncommutativity of coordinates); to look for cross-fertilization among ideas of one branch (say gauge theory models) with those of another branch (say gravity or gravity-like models) trying to fit experimental data.

It is time to introduce in some details the work presented in the last Section where I reprint a collection of 17 papers grouped by areas that well represent my main research activities. This will be done in the following three Sections and Fig.~\ref{scheme} should be of help in understanding the way the papers are grouped and the areas they contribute to. Before doing so I want to mention the research activity I recently added to my traditional ones and that I decided not to include in this Dissertation because it is still premature. That is the investigation of physical problems relevant for biological systems \cite{virustalk}, \cite{cejb}, \cite{dice2008}, \cite{neuro}, \cite{dnaletter}, \cite{dnapaper}, \cite{dnatorus}. This enterprize is undertaken within a theoretical perspective that, on the one hand, pushes till biology the cross-fertilization between branches of physics we mentioned earlier, on the other hand, that hopes to reduce (or enhance...) biology, or sectors of it, to a level of rigor similar to that of theoretical physics (see the invited article \cite{neuro} for more details). Both directions are challenging and highly rewarding.

\section{\sc QFT Vacuum and Quantum Groups}

The vacuum state of an interacting theory in QFT is all but empty. It is actually better viewed as a medium with definite physical properties and it is better called {\it ground state} rather than {\it vacuum}, but here we shall use either terms without warning. To many these properties of the ground state are the most significant experimentally testable effects of QFT (see, e.g., \cite{umezawa2}, \cite{milonni}, \cite{tdlee} among many others): quantum mechanics (QM) alone cannot account for the structure of the ground state, we need QFT. This is a remarkable success of a theory that was born to reconciliate QM and special relativity (SR): in {\it a limit} it gives back the results of QM but it predicts phenomena that QM alone cannot explain.

That in QFT the vacuum state of interacting fields - and consequently the Hilbert space built from it - is an object that deserves a deep mathematical analysis, and that this problem is at the core of the occurrence of the infinities of QFT is the well known subject of many efforts that started with QFT itself \cite{pctspin} and that in the early days focused on the highly relativistic (high energy) nature of fundamental interactions\footnote{This is not well known just to those interested in the foundations and in the mathematically sound formulations of QFT (or Axiomatic QFT), see, e.g., \cite{axiomatic}, but is well known as well to the everyday practitioner that computes cross sections of scattering processes, even if sometimes she/he does not realize that because renormalization procedures are routine applications.} \cite{diracqft}. Nowadays this non-trivial structure of the Hilbert space of QFT is fully appreciated due to the possibility to describe within it statistical/themodynamical aspects of quantum systems.

Applications can be found in fields ranging from condensed matter physics \cite{umezawa1} to thermal physics of black holes \cite{wald} and more (see \cite{vitiello1} for a recent overview). Hence SR may or may not be important. What is important is to have within one large Hilbert space different Fock spaces\footnote{They are sometimes called ``physical spaces'' and some other times, by the mathematically educated, ``superselection sectors'' \cite{haag}.} each of which is not reachable from any other with a transformation that preserves the length of the state vectors, hence the probabilities. These superselection sectors can account for the existence of different phases hence are {\it the quantum frame} to approach spontaneous symmetry breaking phenomena. This is not to say that the matter is fully clarified and that there are not many interesting open questions such as the systematic understanding of the relations between UIRs and divergences in QFT \cite{lectnotes2}, \cite{mills}, or between UIRs and the topologically distinct vacua of Yang-Mills theory \cite{lectnotes3}, \cite{rajaraman}. On the latter is focusing the MSc work of one of my students \cite{derco}.

One way to see this is to consider as starting point the \textit{continuum} nature of classical fields\footnote{Here ``continuous'' is meant in the opposite sense to ``discrete'' and not in the topological sense. Thus in this category fall also function(al)s that indeed have a singularity at the classical level, like the gravitational field of a black hole, the gauge field of a Dirac monopole, etc..}. The field may be localized in a small portion of spacetime\footnote{Examples are the nonabelian gauge fields that mediate the strong and weak interactions, by their intrinsic nature, but also electromagnetic fields that fade away outside a given domain or the finite volume samples of condensed matter: a superconductor, a magnet, a crystal.} or can be over the whole of it or even coincide with it, like the gravitational field, nonetheless it is a continuum that it describes. In QFT the number of degrees of freedom, consistently with the classical case, is defined as the number of annihilation (creation) operators, say $a_k$ ($a^\dagger_k$), necessary to describe the field. This is clearly infinite because $a_k$ and $a^\dagger_k$ appear in the Fourier decomposition of the field and infinitely many plane-waves, say $U_k(t, x)$ are necessary (in 1 space and 1 time dimensions and leaving aside unimportant spin-dependent considerations)
\begin{equation}\label{field}
\phi(t, x) = \int_{- \infty}^{+ \infty} dk \, \left( a_{k}U_k(t, x)\,+\,a_k^{\dagger} U_k^{*}(t, x) \right) \;,
 \end{equation}
where $U_k(t, x)$ obey an orthonormality condition with respect to an inner product like, e.g., the Klein-Gordon one $(\phi_1, \phi_2)=i\int
\phi_1^*\stackrel{\leftrightarrow}{\partial_t}\phi_2 d x$.

Thus the infinity comes in twice: once because there is no $|k_{\rm max}|$, and once more because there are infinite $k$s in any given interval $(k', k'')$, no matter how small. The first infinity might be cured by a fundamental length\footnote{I assume periodic boundary conditions for the field here hence, in a way, I am already assuming finite system size. On the other hand, this size might be taken as large as wanted.} $\ell = 2 \pi / |k_{\rm max}|$ (discretization in configuration space), the second by a finite volume $L = 2 \pi / |k_{\rm min}|$ (discretization in momentum space). Since, e.g., $E = \sqrt{m^2 + k^2}$, the first infinity is reached at the cost of infinite energy - i.e. the energy necessary to have quanta interacting at arbitrarily small separation distances - the second infinity, instead, is reached at zero energy cost and is related to the vacuum field configurations attained at $L \to \infty$ (the $V$-limit).

An infinity of {\it states}, say $|n \rangle$, $n = 1, 2, ...$, can also appear in QM, so that the Hilbert space associated with the given system can be infinite dimensional, but it is always {\it separable}\footnote{In the language of set theory: A Hilbert space is separable if it contains a countable dense subset. A set is said to be infinite countable if its cardinality, $\aleph$, is the same as that of the natural numbers, $\aleph_0$. The cardinality of a set cannot be changed by a {\it finite} number of operations.}. For instance, the energy eigenstates of one harmonic oscillator are the Hermite polynomials, $\langle x | n \rangle = H_n(x)$, but the degree of freedom is only one, $a$ ($a^\dagger$). This infinity is of a similar nature as the first infinity of the field theory in the sense that it is only attained at the price of infinite energy, $E = (n + 1/2) \hbar \omega \to \infty$ for $n \to \infty$.

The Hilbert space associated with the field in (\ref{field}) is ${\cal H} =  {\cal H}_1 \bigotimes {\cal H}_2 \cdots \bigotimes  {\cal H}_N$, where $N = L / \ell = |k_{\rm max}| / |k_{\rm min}|$ is the number of degrees of freedom. Each ${\cal H}_i$ is infinite dimensional ${\cal H}_i = \{ | n_i \rangle \}$, $n_i = 0, 1, 2, ...$, just like for the harmonic oscillator, and until $N$ is finite we can use the fact that the Cartesian product of finitely many countable sets is countable (and that tensor products are built-up out from Cartesian products). For instance, we could apply the Cantor pairing to the first two members of the family and map $\{ | n , m \rangle \}$ to the natural numbers, and then go on by induction for all members of the family till $N$. The key point is that when $N$ is infinite $\cal H$ is no longer a infinite countable set and when we want to select out of it an infinite countable set (the ``physical space'') there are infinitely many inequivalent ways of doing it, the reason being clarified formally in the von Neumann theorem \cite{vNeumann}.

This being an old and very important problem excellent books have been dedicated to its rigorous mathematical study (see, e.g., the classic works \cite{pctspin}, \cite{axiomatic}, \cite{haag}) and to its physical consequences (see, e.g., \cite{umezawa2}, \cite{umezawa1}). For a textbook discussion see, e.g., \cite{schweber}. For a recent PhD dissertation see \cite{tracy}. I dedicated part of my Laurea (MSc) thesis \cite{thesis1} and my first published paper {\bf \cite{qVN}} (included in Section~\ref{selpap}) to the von Neumann theorem. There I present rigorous proof and discussions linking on the general ground the UIRs and the q-deformed algebras of canonical (second) quantization. Therefore my expertise with the problem is hopefully shown there. Here I would like to give a simple and intuitive proof, only valid for a particular case, that will help me introduce the condensate structure of the QFT vacuum and the related statistical/thermodynamical formalism that is used, directly or indirectly, in the applications (via q-groups) to quantum dissipation {\bf \cite{qQD}}, to the algebraic structure of ThermoField Dynamics (TFD) {\bf \cite{pla}} and to thermal properties of black holes {\bf \cite{qBH1}} that I present in the collected papers.

Suppose we have a complex version of (\ref{field}), with $N$ degrees of freedom
\begin{equation}\label{complexfield}
 \phi(x)=\sum_{k = - N/2}^{N/2} \, \left( a_{k}U_k(t, x) \,+\, b_k^{\dagger}U_k^{*}(t, x) \right) \;,
\end{equation}
where
\begin{equation}\label{algebraab}
[a_k, a_{k'}^\dag]=\delta_{kk'}\,, [b_k, b_{k'}^\dag]=\delta_{kk'} \,, \quad [a_{k} , b_{k'}] = [a_k, b^\dagger_{k'}] = 0 ,
\end{equation}
and
\begin{equation}
a_k |0\rangle = b_k |0\rangle = 0 \;.
\end{equation}
The Hilbert space, say ${\cal H}(a,b)$, is constructed by repeated actions of the creation operators $a_{k}^\dag$ and $b_k^\dag$ on $|0\rangle$ and is infinite dimensional but separable. The Bogolubov transformations are those that leave (\ref{algebraab}) invariant and are the
hyperbolic $SU(1,1)$ transformations
\begin{eqnarray}
c_k & = & a_k \cosh \theta_k - b_{- k}^\dagger \sinh \theta_k = G^{-1}(\theta) a_k G(\theta) \,, \label{bog1} \\
d_k & = & b_k \cosh \theta_k - a_{- k}^\dagger \sinh \theta_k = G^{-1}(\theta) b_k G(\theta) \,, \label{bog2} \,,
\end{eqnarray}
with
\begin{equation}\label{algebracd}
[c_k, c_{k'}^\dag]=\delta_{kk'}\,, [d_k, d_{k'}^\dag]=\delta_{kk'} \,, \quad [c_{k} , d_{k'}] = [c_k, d^\dagger_{k'}] = 0 ,
\end{equation}
and
\begin{equation}\label{eq16}
G(\theta) = \exp \left\{  \sum_{k = - N/2}^{N/2} \theta_k \left[ a_k b_{-k} - b_{- k}^\dagger a_k^\dagger \right] \right\} \,{.}
 \end{equation}
The Hilbert space, say ${\cal H}(c,d)$, is constructed as before by repeated actions of $c_{k}^\dag$ and $d_k^\dag$ on $|O\rangle$, where
\begin{equation}
c_k |O\rangle = d_k |O\rangle = 0 \;,
\end{equation}
and, since $G(\theta)$ in (\ref{eq16}) is a unitary operator $G^{-1}(\theta)=G(-\theta)=G^{\dagger}(\theta)$, ${\cal H}(c,d)$ is just the
same space as ${\cal H}(a,b)$
\begin{equation}\label{Ovso}
|O\rangle \, = \,G^{-1}(\theta)|0 \rangle \,.
 \end{equation}
By using the Gaussian decomposition for $G$ this vacuum can be expressed as a $SU(1,1)$ generalized
coherent state \cite{perelomov} of Cooper-like pairs
\begin{equation}\label{coherent}
|O \rangle = f (\theta) \,\exp\left\{ \sum_{k = - N/2}^{N/2} \;\tanh\theta_k a_k^\dagger b_{- k}^\dagger \right\} \, |0 \rangle\,{,}
\end{equation}
which is the condensate structure we sought for. Here
\begin{equation}
f(\theta) = \prod_{k = - N/2}^{N/2} \; (\cosh\theta_k)^{-1} \;.
\end{equation}

Now let me first set $\theta_k \equiv \theta =$ constant for all $k$s and then let me compute the product of the vacuum states
\begin{equation}
\langle 0 | O \rangle = f (\theta) = (\cosh \theta)^{-N} \;.
\end{equation}
For finite $N$ this product is finite, but for infinite $N$, recalling that $\cosh \theta > 1$ for $\theta \neq 0$,
\begin{equation}
\langle 0 | O \rangle \to 0 \quad {\rm for} \quad N \to \infty \;.
\end{equation}
This is the simplified version of the von Neumann theorem I was seeking for\footnote{The same conclusions would be reached in the case of a fermionic field. What changes is the compactness of the Bogolubov transformations (they are $SU(2)$ rather than $SU(1,1)$) hence the associated Gaussian decomposition for $G$. Eventually, with the simplifying assumption $\theta_k \equiv \theta$, $f(\theta) = (\cos \theta)^{N}$ and clearly $f(\theta) \to 0$ for $N \to \infty$.}: when $N$ is finite we are in the hypothesis of the theorem, hence every transformation of the kind (\ref{bog1}) and (\ref{bog2}) is a mere change of basis within the same Hilbert space (the spaces are unitarily equivalent); when $N$ is infinite the Hilbert space associated to the transformed creation and annihilation operators is {\it orthogonal} to the Hilbert space associated to the untransformed creation and annihilation operators. The spaces now are said to be unitarily inequivalent and (\ref{algebraab}) and (\ref{algebracd}) are unitarily inequivalent representations of the canonical commutation relations. The rigorous language is in {\bf \cite{qVN}}. In my papers {\bf \cite{qQD}}, {\bf \cite{qBH1}} and {\bf \cite{TFDBH}} presented in Section~5 a detailed discussion is given on how to handle this situation, on how to perform the limit $N \to \infty$ and on how to construct the vacuum $|O \rangle$ in relation to quantum dissipation, quantum algebras and certain properties of QFT in the presence of an event horizon.

The limit $N \to \infty$ may not be attained in real systems, nonetheless, QFT does not have a natural cut-off and the infinite volume
($L$ for us here) limit is customary to include zero energy modes. On the other hand, even a more refined discussion that includes, e.g.,
gravitational effects, such 't~Hooft-Susskind's holographic principle \cite{thooftholography}, hence the ``natural'' Planck scale for $\ell$ ($\ell_P \sim 1.6 \times 10^{-35}$m), would give $N \simeq 10^{32}$ for, e.g., a ``sample'' of one millimeter, leading to
\begin{equation}
\langle 0 | O \rangle \sim (\cosh \theta)^{- 10^{32}} \;,
\end{equation}
a number that could be taken as zero with confidence.

Here $\phi (x)$ is a free field and this means that the UIRs are a mathematical fact due to the infinite $N$. Nonetheless, the physical relevance of UIRs only appears when the interaction is present because it is in that case that time evolution becomes impossible to describe within a single Hilbert space \cite{diracqft}.

The condensate structure of (\ref{coherent}) is the starting point to study thermal/statistical properties of various systems. As briefly mentioned, for instance, the vacuum for a quantum dissipative system can be seen as a condensate of quanta describing the system and the environment {\bf \cite{qQD}}; the vacuum of TFD (where the reservoir degrees of freedom, or tilde fields, are introduced from the very beginning) has a similar structure {\bf \cite{pla}}; some general features have been investigated in {\bf \cite{qBH1}} and \cite{qBH2}, with the emphasis on the role of the gravitational (classical) back-ground\footnote{Incidentally, the link between UIRs and TFD in black hole physics (through the seminal work of Israel \cite{israel}) and the pivotal Maldacena's conjecture of a gravity/gauge theory correspondence \cite{maldacena} was depicted in \cite{maldacena2}.}.

Such thermal/statistical properties are made evident when, through algebraic manipulations, the vacuum (\ref{coherent}) is written in terms of an entropy operator, or of a coherent states/squeezing operator, etc. Some of those re-writings are reviewed in \cite{thesis1} and are illustrated within different contexts in {\bf \cite{qQD}}, \cite{qTFDproc}, {\bf \cite{qBH1}}, \cite{qBH2},  \cite{qBHproc2}, \cite{qBHproc1} and references therein. The range of the applications of these and related techniques is wide. For instance, one application we contributed to start-up is that to neutrino oscillations {\bf \cite{neutrino}}, \cite{neutrinoapb} (where the mathematical tool of q-groups was not employed), a line of research that has grown over the years (see, e.g., \cite{capolupo} for a review), and in the collected works we also present an application of the methods of coherent states to the study of a quantum system of interest for quantum computing, namely the system that should implement the Shor factorization algorithm {\bf \cite{shor}} (a line of research that, through various stages, brought me and my coworkers to explore the role of quantum physics in biology \cite{dice2008}, in particular that of QFT for DNA interactions \cite{dnaletter}, \cite{dnapaper}). It is time, though, to briefly introduce the main mathematical tool used in most of the included papers on this topic, q-groups, and to explain why it is relevant.

In a nutshell and in simple words, in \cite{thesis1} we proved (see the published work {\bf \cite{qVN}} and also \cite{banff}) that the parameter labeling the UIRs of QFT is the parameter $q$ that characterizes a generalized version of (\ref{algebraab}). This is so because the generator of the Bogolubov transformations is (up to a c-number) the generalized commutator $[a_q, \hat{a}_q]$, where $a_q \to a$ and $\hat{a}_q \to a^\dag$ for $q \to 1$. With this result in hands and after having explained the role of UIRs and of Bogolubov transformations, it should be evident that such generalized versions (or deformations) of the canonical algebra find many applications where the UIRs are relevant. It is some of these applications that we explored in \cite{thesis1}-{\bf \cite{pla}}, {\bf \cite{qBH1}}, \cite{qBH2}, \cite{qBHproc2}.

Let me now make more explicit some of the ideas behind q-groups (for a brief and delightful introduction to q-groups see \cite{majid1}, for a textbook introduction see, e.g., \cite{majid2}). I shall do that in a different spirit as that of \cite{thesis1}, \cite{banff}, {\bf \cite{qVN}} which means that from the following it will not be clear why $[a_q, \hat{a}_q]$ behaves like the generator of the Bogolubov transformations (for this is well explained in the presented paper {\bf \cite{qVN}}) but it would be clear the role of the bi-algebra structure that is essential for the application to quantum dissipative systems {\bf \cite{qQD}}, to TFD  \cite{qTFDproc}, {\bf \cite{pla}}, black hole physics {\bf \cite{qBH1}}, \cite{qBH2}, \cite{qBHproc2}, \cite{qBHproc1}, etc.

Consider a single quantum degree of freedom, $a, \, a^\dagger$. On a very general ground  $(a, \, a^\dagger, \, {\cal N}, \, c )$ form what is called the {\it Weyl-Heisenberg algebra} $h(1)$ if
\begin{equation}\label{a}
 [a,a^{\dagger}] = 2c\,, \quad [{\cal N},a] = -a\,, \quad  [{\cal N},a^{\dagger}]
 = a^{\dagger}\,, \quad [c, \cdot] = 0 \;.
 \end{equation}
In the fundamental representation, with central term $c \equiv \frac{1}{2}$ and Casimir $C = 2c N - a^\dagger a \equiv 0$ , this algebra reduces to the standard algebra of quantization and ${\cal N} = a^\dagger a$.

To the algebra $(a, \, a^\dagger, \, {\cal N}, \, c )$ it is natural to associate a {\it Hopf algebra}, i.e. its {\it universal enveloping algebra}\footnote{The universal enveloping algebra of a Lie algebra $\bf g$ is customarily denoted by ${\cal U} ({\bf g})$, and $h(1)$ makes no exception. On the other hand, not to clutter the formulae and due to the small use of the enveloping algebra I shall make here, with the exception of the coproduct operation $\Delta$ (see later), I shall employ the same symbol for both the algebra and the universal enveloping algebra.} \cite{majid2}. In there, besides the ordinary multiplication of (\ref{a}), there are three more operations: the \textit{coproduct}, the \textit{counit} and the \textit{antipode} \cite{majid1}, \cite{majid2}. What I want to emphasize here is the role of coproduct $\Delta$
\begin{equation}\label{co}
\Delta a = a \otimes I + I \otimes a \;,
\end{equation}
(similarly for $a^{\dagger}, \, {\cal N}$ and $c$) that is a homomorphism, and whose physical meaning is that it provides the prescription for operating on two modes, a crucial fact for the applications discussed in \cite{qTFDproc}, {\bf \cite{pla}}, {\bf \cite{qBH1}}, \cite{qBH2}, \cite{qBHproc2}.

It is worth noting that such an operation is not as exotic as it might look at first glance. For instance, the familiar operation of addition of the angular momentum $J^i \in su(2)$, $i = 1,2,3$, of two particles is a coproduct
\begin{equation}
 \Delta J^i = J^i \otimes I + I \otimes J^i = J_1^i +J_2^i \,.
\end{equation}
In other words, the natural assumption of the additivity of basic observables, such as the energy and the angular momentum, necessarily implies to consider the coproduct operation, namely the Hopf algebra structure \cite{qTFDproc}, \cite{qBHproc2}, \cite{qBHproc1}.

The deformation of $h (1)$, denoted by $h_q (1)$,  is given by
 \begin{equation}\label{aq}
[a_q,a_q^{\dagger}] = [2c]_q\,, \quad
[{\cal N}, a_q] = -a_q\,, \quad [{\cal N},a_q^{\dagger}] = a_q^{\dagger}\,, \quad [c, \cdot] = 0 \;,
 \end{equation}
where ${\cal N}$ and $c$ are the same as in $h(1)$ and \textit{primitive} (i.e. with coproduct given by (\ref{co})), the commutator of $a_q$ and $a_q^{\dagger}$ is different from the commutator of $a$ and $a^\dagger$ for generic $c$, and $[x]_q \equiv \displaystyle{\frac{q^x - q^{-x}}{q - q^{-1}}}$. Hence for the coproduct of $a_q$ we obtain
\begin{equation}\label{coq}
\Delta a_q = a_q \otimes q^{c} + q^{-c} \otimes a_q \,,
\end{equation}
where I use the property $[\Delta a_q, \Delta a_q^\dagger]=\Delta([a_q, a_q^\dagger])$, and similarly for $a^\dag_q$. The parameter $q$ is called the
{\it deformation parameter}.

For $c=\frac{1}{2}$ ($[1]_q = 1$) the algebra (\ref{a}) reduces to the standard CCRs of quantization, and the only difference between $h(1)$ and $h_q (1)$ is in the coproduct. The two algebras act on the same representation space, the Hilbert-Fock space $\cal H$, and both $\Delta$ and $\Delta_q$ map operators acting on $\cal H$ to operators acting on $\cal H \otimes \cal H$. For the applications reported here the key step is to consider the algebraic structure of the entire field, i.e. to take into account the infinite number of degrees of freedom. Let me focus for instance on one set of the operators of (\ref{complexfield}) and (\ref{algebraab}), for instance $(a_k, \, a_k^\dag, \, {\cal N}_k )$, as for the set relative to the antiparticles everything is the same. In this case we have an infinite number of copies of the algebra $h(1)$, one for each $k$,
\begin{equation}\label{ak}
[a_k,a_k^{\dagger}] = 1\,, \quad [{\cal N}_k,a_k] = -a_k\,, \quad
[{\cal N}_k,a_k^{\dagger}] = a_k^{\dagger}\,, \qquad \forall k\,,
\end{equation}
and different copies $h_k (1)$ and $h_{k'} (1)$, $k \ne k'$, are related by the standard commutation relations for a quantized field
 \begin{equation}\label{aa'}
 [a_k,a_{k'}^{\dagger}] = \delta_{k k'}\,,  \quad [a_k,a_{k'}] =
 [a_k^{\dagger},a_{k'}^{\dagger}] = 0\,, \qquad \forall k, k'\,.
 \end{equation}
The deformation of each of the $h_k (1)$s is exactly the same as for the single mode. We have then just an infinite set of deformed
algebras labeled by $k$, and if we set $ h_k (1) \rightarrow h_{q(k)} (1) $ we have
\begin{equation}\label{aa'q}
[a_{q(k)} , a_{q(k')}^\dag] = [\delta_{k k'}]_{q(k)} = \delta_{k k'}\,, \qquad \forall k, k'\,,
\end{equation}
similarly for the other commutators. More generally, the deformation parameter $q$ could depend on a momentum $p$ which may or may not coincide with $k$.

It should now be clear that two are the things that make q-groups natural structures to consider when dealing with UIRs and the related statistical/thermal properties. On the one hand, the deformation parameter is not an external parameter but (an important) part of the algebraic structure of the quantum field\footnote{$q$ is the label of the UIRs. See {\bf \cite{qVN}}.}, and it could be made to depend also on other parameters: $q=q($phys$,p)$. Here ``phys'' stands for the physical quantity appropriate for the given system under study: damping constant (for the application to quantum dissipation {\bf \cite{qQD}}), temperature (for the application to TFD  \cite{qTFDproc}, {\bf \cite{pla}}), surface gravity (for the application to Hawking radiation {\bf \cite{qBH1}}), acceleration (for the application to Unruh radiation {\bf \cite{qBH1}}), etc. On the other hand, the bialgebra structure, hence the coproduct, are pivotal to handle the doubling of the degrees of freedom typical of a system in thermal contact with a reservoir, a situation common to all the cases just mentioned: a dissipative system needs a reservoir into which to dissipate; TFD is constructed via the doubling of the degrees of freedom in the first place; the Hawking and Unruh effects are based on the existence of two causally disconnected (but quantum entangled) regions of space-time.

\section{\sc Spatiotemporal Symmetries}\label{stsymm}


The importance of Poincar\'e symmetry for physics was established a bit more that one hundred years ago, it is natural then that many attempts have been made to go beyond it, one way or the other. Conformal and Super symmetries are one way to go beyond Poincar\'e symmetry by the road of generalizing it to larger symmetry (super-) groups that include Poincar\'e as a particular case. Noncommutativity of spatiotemporal coordinates (at least in a setting) is one way to go beyond it by invoking Lorentz symmetry violation.

Both these approaches had an enormous impact on mayor research and a conspicuous part of their motivations lays in the search for physics beyond QFT and in the search for a meaningful theory of quantum gravity, and none of them is still there \cite{3qs}. On the other hand those ideas (as often happens to deep and mathematically sound structures) find applications in or are realized by a variety of physical systems that are within the reach of our experiments, hence are proving to be useful for real physics.

\subsection{\sc Conformal and Super Symmetries}\label{confsusy}


When the spatiotemporal coordinates undergo an infinitesimal change
\begin{equation}
x^\mu \to {x^{\prime}}^\mu = x^\mu - f^\mu (x) \;,
\end{equation}
the fields, say $\Phi_i (x)$, respond according to their spin
\begin{eqnarray} \label{trnfsPhi}
\delta \phi = 0 \;, \; \delta V_\mu = (\partial_\mu f^\nu) V_\nu \;, \; \delta V^\mu = - (\partial_\nu f^\mu) V^\nu \;, \; {\rm etc.}
\end{eqnarray}
for $\Phi_i$ scalar, covariant vector, contravariant vector, respectively, where
$\delta \Phi_i \equiv \Phi^{\prime}_i (x') - \Phi_i (x)$. By introducing $\delta^* \Phi_i \equiv \Phi^{\prime}_i (x) - \Phi_i (x)$,
i.e. the response to the coordinate change evaluated at the same point\footnote{See, e.g., pgs 243-245 of \cite{jackiwbook}.}, we see that
\begin{equation}\label{geomtrnfs}
\delta^* \Phi_i = f^\mu \partial_\mu \Phi_i + \delta \Phi_i  \;,
\end{equation}
hence from (\ref{trnfsPhi}) and from the definition of the Lie derivative
\begin{eqnarray}
{\bf L}_f X_{\mu \dots \nu}^{\lambda \dots \kappa} &=& f^\alpha
\partial_\alpha X_{\mu \dots \nu}^{\lambda \dots \kappa} + (\partial_\mu
f^\alpha) X_{\alpha \dots \nu}^{\lambda \dots \kappa} + \cdots +
(\partial_\nu f^\alpha) X_{\mu \dots \alpha}^{\lambda \dots
\kappa} \nonumber \\
&-&(\partial_\alpha f^\lambda) X_{\mu \dots \nu}^{\alpha \dots
\kappa} - \cdots - (\partial_\alpha f^\kappa) X_{\mu \dots
\nu}^{\lambda \dots \alpha} \;, \label{lie}
\end{eqnarray}
one sees that\footnote{\label{notefermions}What I just said is true only for integer spin fields. E.g., in the case of tensor fields
and for the $f$s of Lorentz transformations (see (\ref{6})), the spin-connection contribution is fully taken into account by the
terms that include derivatives of $f$, as can be easily read-off from (\ref{lie}). For half-integer spin fields,
like Majorana, Weyl, Dirac, Rarita-Schwinger, etc. some more work is needed to explicitly introduce the spin-connection. Nonetheless,
keeping this warning in mind, for the purpose of  introducing the conformal transformations in a compact and
geometrically sound fashion, what I do here should lead to no confusion as I eventually shall exhibit just the
infinitesimal action on a scalar field, see (\ref{phitranls})-(\ref{phispecconf}).}
\begin{equation}
\delta^* \Phi_i = {\bf L}_f \Phi_i \;,
\end{equation}
and all spatiotemporal transformations can be treated in a compact way \cite{confLie}
\begin{equation}
[ {\bf L}_f , {\bf L}_g ] \Phi_i = {\bf L}_{[f,g]} \Phi_i \;,
\end{equation}
where standard notation is employed {\bf \cite{NCNoether}}, \cite{lectnotes3}. The latter expression is at once
geometrical (as it refers to Lie derivatives of fields) and algebraic (as the group structure of the
given transformations can be easily made explicit).

We are not saying anything yet on whether the $f$s can be symmetries of a physically sound theory described by an action built out from the fields and their derivatives, as it would be necessary for the application of the Noether theorem to extract conserved physical quantities like energy, momentum, angular momentum, etc. (for a general discussion with the notation introduced here see {\bf \cite{NCNoether}}, {\bf \cite{NCNother2}}, \cite{lectnotes1},
\cite{lectnotes2}). Indeed, the level at which we are moving is that of general coordinates transformations and {\it all} spatiotemporal transformations are included in this description. We want to identify, though, the most general coordinates transformations that can be physical symmetries in flat space. We can do so by requiring the transformations to preserve the light-cone structure of space-time
\begin{equation}\label{Lgsg}
{\bf L}_f g_{\mu \nu} = \sigma g_{\mu \nu} \;,
\end{equation}
and then consider the limit $g_{\mu \nu} \to \eta_{\mu \nu}$.
This is the most fundamental/less demanding request we can do on transformations to be symmetries of physical
theories, i.e. that they respect causality at a geometrical level. Indeed there are physical theories like Maxwell
electrodynamics, or the classical theory of color interaction for hadrons, that are not sensitive to a change in scale
of the observation. Of course, this is immediately recognized as an approximation that at some stage needs to fail,
even just for the fact that a natural scale for all physical systems is not to be posited but it is given by the
Planck constant $\hbar$. As a matter of fact, is well known that invariance under such transformations is broken when
quantum effects become important, a phenomenon known as conformal anomaly. Nonetheless, to a certain extent
and in a certain limit, the angle-preserving $f$s of (\ref{Lgsg}) are real symmetries of real physical systems
and, interestingly, they are as much as a physical system can sustain for what matters spatiotemporal
transformations: there is nothing more.

I have introduced a mathematical tool and should make use of it to make my statements more precise and see what
is the group structure for this symmetry transformations and how they are related to the more familiar and
``sacred'' (as opposed to ``violated'') Poincar\'e symmetry. All one needs to do is to rewrite (\ref{Lgsg}) in an
appropriate way in, say, $n$ dimensions
\begin{equation}\label{confkill}
\nabla_\mu f_\nu + \nabla_\nu f_\mu  = \frac{2}{n} g_{\mu \nu} \nabla_\alpha f^\alpha \;,
\end{equation}
with $\nabla_\mu$ the covariant derivative compatible with $g_{\mu \nu}$. This is the {\it conformal} Killing equation,
whose only solutions in {\it flat space} are
\begin{equation}  \label{6}
f^\mu = a^\mu \quad {\rm or} \quad f^\mu = \omega^\mu_\nu x^\nu \quad {\rm or%
} \quad f^\mu = a x^\mu \quad {\rm or} \quad f^\mu = a^\mu x^2 - 2
a \cdot x x^\mu \;,
\end{equation}
i.e. infinitesimal translations, rotations/boosts, dilations, and special conformal transformations, respectively, where, as usual, $%
\omega^{\mu \nu} = - \omega^{\nu \mu}$. These form the group $SO(n,2)$ as can be checked by considering the action of the Lie derivative on a scalar field\footnote{Higher spin only matters for more involved issues, see footnote \ref{notefermions} and, e.g., {\bf \cite{weyl}}.} with the $f$s in (\ref{6})
\begin{eqnarray}
{\bf L}_f \phi & = & a^\mu \partial_\mu \phi \equiv  a^\mu P_\mu \phi \;, \label{phitranls}\\
{\bf L}_f \phi & = & \frac{1}{2} \omega^{\mu \nu} ( x_\mu \partial_\nu - x_\nu \partial_\mu) \phi \equiv
\frac{1}{2} \omega^{\mu \nu} M_{\mu \nu} \phi \;, \\
{\bf L}_f \phi & = & a x^\mu \partial_\mu \phi \equiv a D \phi \;, \\
{\bf L}_f \phi & = & a^\mu ( x^2 \partial_\mu - 2 x_\mu x^\nu \partial_\nu) \phi \equiv a^\mu K_\mu \phi \;, \label{phispecconf}
\end{eqnarray}
and by computing the various commutators of $P_\mu, M_{\mu \nu}, D, K_\mu$. This is called the conformal group.
One then sees that $P_\mu, M_{\mu \nu}$ generate the Poincar\'e group, $ISO(n-1,1)$, while with the other two sets of generators, $D$ and $K_\mu$, the commutations relations of $SO(n,2)$ are obtained (see, e.g., \cite{lectnotes3}). The subgroup $ISO(n-1,1)$ is also obtained when the Killing equation
\begin{equation}
\nabla_\mu f_\nu + \nabla_\nu f_\mu  = 0 \;,
\end{equation}
descending from
\begin{equation}\label{Lg0}
{\bf L}_f g_{\mu \nu} = 0 \;,
\end{equation}
is considered instead of (\ref{confkill}). The latter has $f^\mu = a^\mu$ or $f^\mu = \omega^\mu_\nu x^\nu$ as only solutions in flat space. It is clear from (\ref{Lgsg}) and (\ref{Lg0}) in which sense conformal transformations are an enlargement of the Poincar\'e transformations.

As briefly mentioned this fundamental idea finds many applications in physics, ranging from classical (unbroken) gauge theories like Maxwell electrodynamics and color interaction, to the massless limit of the low energy Hamiltonian of graphene\footnote{We are also exploring this direction in the work in progress \cite{weylgraphene}.} \cite{graphenereview}, from systems at the phase transition \cite{difrancesco} to certain gravitational theories existing in three dimensions (two space, one time) \cite{djt}, \cite{hornewitten}. To the latter application I dedicated some of my own work {\bf \cite{cs3}}, \cite{cs3proc} and have some work in progress \cite{gaugevirus}, see also \cite{lectnotes3}.

Conformal symmetry becomes particularly stringent in two dimensions because, as it is easy to see, the $f$s satisfying (\ref{confkill}) are all the holomorphic (and anti-holomorphic) complex functions, hence admit a Laurent expansion and an infinite number of generators, one for each term of the series. These infinite dimensional algebras, called Virasoro algebras \cite{difrancesco} impose so many constraints that the system that carries such a symmetry is usually integrable, in principle. For instance, in the quantum theory, where some care is necessary to handle the conformal anomaly above mentioned, one has Ward-like identities that should determine all N-point correlators\footnote{While it is true that infinite constraints should give rise to fully solvable models, in practice the equations that the Green's functions have to satisfy can be very difficult to solve. The typical example are the Knizhnik-Zamolodchikov equations for the Wess-Zumino-Witten model.}.

The aware reader might have recognized that the way conformal transformations were introduced, via the geometrical transformations (\ref{geomtrnfs}) and through the request (\ref{Lgsg}), is not conventional. Actually what we have introduced in (\ref{Lgsg}) are {\it Weyl transformations} in curved space, whose relationship with scale and conformal transformations in flat space was fully elucidated in the work {\bf \cite{weyl}} presented here. Besides finding there the general criterion to have conformal invariance from scale invariance in flat space for theories of fields of arbitrary spin in arbitrary dimensions, in that work the intriguing issue of the ``improvements'' for the energy momentum tensor was clarified in a systematic way. Improvement terms are boundary terms that are added to the energy momentum tensor associated to invariance under translations (canonical tensor) in order to have symmetry under the rest of the conformal group, i.e. to make it symmetric and traceless\footnote{It is the amount by which the energy-momentum tensor fails to be traceless in the quantum theory that is called the conformal or scale anomaly and, in two dimensions, it is related to the central term of the Virasoro algebra we briefly introduced and it involves $\hbar$, see {\bf \cite{weyl}}.}, see, e.g., \cite{ccj}. The results established in {\bf \cite{weyl}} have been used by practitioners in the field in various general theoretical investigations, see e.g. \cite{weyljackiw} for a recent investigation related to two dimensions. One direction we just started pursuing is the application of the above outlined results to the conformally invariant limit of the effective Hamiltonian for the low energy theory of graphene \cite{weylgraphene}.

A way of looking at improvement terms is to see them as a signature of spatiotemporal symmetries as opposed to {\it internal} symmetries (i.e. symmetries based on transformations only involving fields' degrees of freedom, such as isospin) for which no improvement is necessary to obtain the conserved charges. This is one of the subtle differences between spatiotemporal and internal symmetries that were explored in the 1960's in the
attempt to combine Poincar\'e invariance and internal symmetry (see, e.g., my lectures \cite{lectnotes1}, \cite{lectnotes2} and the review \cite{nogo} and references therein). That effort produced various ``no-go'' theorems, the Coleman-Mandula's and the O'Raifeartaigh's being the most important, see \cite{nogo} for a review. To say it in a nut-shell, those theorems establish, to various degrees of generality, that if  $E$ is the symmetry Lie group of the theory (e.g., of the $S$ matrix), containing the Poincar\'e group, say $L$, as subgroup, and if some natural assumptions hold, then $E$ can only be the direct product of the Poincar\'e group and the internal symmetry group $T$, $E =  L \times T$ \cite{nogo}. This clearly is an attempt to go beyond Poincar\`e invariance. It was pursued in the 1960's due to the phenomenological successes of the $SU(6)$ model of strong interaction to explain the static quark model based on three flavors only. That model failed but the general results of the ``no-go'' theorems are solid pillars on which yet another attempt to generalize Poincar\'e symmetry has been based. Such attempt is SUSY, a symmetry that indeed succeeds in combining non trivially (i.e. not as a direct product) internal and spatiotemporal symmetries. I shall now sketch the basic ideas behind it to introduce some of my work in the field \cite{thesis2}, {\bf \cite{swplb}}, {\bf \cite{swannals}}, {\bf \cite{sucs}}, see also \cite{lectnotes1}. Before doing so a warning is necessary. It is only speculated that SUSY is (or was) a fact of nature, at least in the spirit of being a fundamental symmetry hence of predicting a SUSY partner for every fundamental particle known to date. On the other hand, some aspects of it find applications in disparate areas of physics, such as certain quantum mechanical models \cite{susyqm}, nuclear physics \cite{iachello} and DNA physics \cite{susydna}.

The key objects of SUSY are certain spin-$1/2$ charges\footnote{The SUSY conventions I use are those of Wess and Bagger \cite{wb}.
Spinors are Weyl two components in Van der Waerden notation, i.e. undotted indices transform under the representation
$(\frac{1}{2},0)$ of $SL(2,\C)$, the covering of the homogeneous Lorentz group, while dotted indices transform under the conjugate
representation $(0,\frac{1}{2})$. The relations between Dirac, Majorana and Weyl spinors are given by
\[
\Psi_{\rm Dirac} = \left(
\begin{array}{c} \psi_\alpha \\  \blambda^{\dalpha} \end{array}
\right) \quad \Psi_{\rm Majorana} = \left( \begin{array}{c}
\psi_\alpha \\  \bpsi^{\dalpha} \end{array} \right)\;.
\]
To raise and lower spinor indices use $\eps_{\alpha \beta}$ and $\eps^{\alpha
\beta}$, where $\eps_{2 1} = \eps^{1 2} = - \eps_{1 2} = 1$.}, $Q^L_\alpha$, that have the following commutation and anti-commutation relations with Poincar\'e generators, internal symmetry generators and among themselves
\begin{eqnarray}
{[} Q_\alpha^L , \bar{Q}_{\dalpha M} ]_{+} &=& 2
\sigma^\mu_{\alpha \dalpha} P_\mu \delta^L_M  \label{qqbar}\\
{[} Q_{\alpha}^L , Q_{\beta}^M ]_{+} &=& \eps_{\alpha \beta} Z^{L M}  \label{qq} \\
{[} Q_{\alpha}^L ,  P_\mu ]_{-} &=& 0 \label{qp} \\
{[} Q_{\alpha}^L ,  M^{\mu \nu} ]_{-} &=& \sigma^{\mu \nu \; \beta}_{\;\;\; \alpha} Q_{\beta}^L \label{qm} \\
{[} Q_\alpha^L , T_l ]_{-} &=& (S_l)^L_M Q_\alpha^M \;. \label{qt}
\end{eqnarray}
Here the $T_l$s are the internal symmetry generators obeying ${[} T_l , T_m ]_{-} = i C_{l m}^k T_k$, the $Z^{L M} = (a^l)^{LM} T_l$ are central terms (i.e. they commute with all other generators and among themselves, hence belong to the Abelian sector of the internal symmetry algebra) and the coefficients $(S_l)^L_M$ obey the same algebra as the $T_l$s and are related to the $(a^l)^{LM}$s via $(S_l)^L_{\; \; M}  (a^m)^{M K} = - (a^m)^{L M} (S^{* l})_M^{\; \; K}$. The upper-case Latin indices run from 1 to N to take into account the so-called N-extend SUSY algebra, which is the most general form SUSY can have. In theories where SUSY holds the $Q^L_\alpha$s are conserved due to the effects of the Noether theorem\footnote{\label{susynote} There are many subtleties regarding the application of Noether theorem to SUSY Lagrangians \cite{thesis2}, \cite{lectnotes1}, see also \cite{susyweinberg}. A recipe to deal with these and other difficulties was one of the outcomes of my own work reported here {\bf \cite{swannals}}.}.

It is clear from (\ref{qqbar}) that SUSY is a spatiotemporal symmetry (in a way it is the ``square root'' of $P_\mu$) and from (\ref{qq}) and (\ref{qt}) that indeed it has non trivial commutations with the internal symmetry. This is made possible by the assumption that spin-$1/2$, hence fermionic, generators are included, an instance not considered by the ``no-go`` theorems that has striking implications, the most immediate being that the fermionic and bosonic degrees of freedom of SUSY models need to be the same because the transformations generated by $Q_\alpha^L$ act on the spin of the fields changing it by $1/2$. This means that a prototypical multiplet under this symmetry is\footnote{This is the $N=1$ on-shell chiral multiplet of the Wess-Zumino model, the first four-dimensional SUSY field theory model discovered.} $(\phi, \psi_\alpha)$, and SUSY does not distinguish among the component fields. If SUSY is taken as a fundamental symmetry that generalizes Poincar\'e or conformal symmetry, then every fundamental particle discovered to-date should have a SUSY counterpart, and this is not seen in experiments.

Thus, about forty years after its proposal, the relevance for physics of SUSY as a fundamental symmetry of nature is still a fully open question. This did not stop serious research in the field which, on the contrary, along with ST (the tightly related ``sister topic'' to SUSY) became the mayor concern of the high energy theoretical community (see, e.g., \cite{3qs} for an invitation to a critical view).

Keeping the above in mind, SUSY in particle physics has many features that make it very appealing to the theoretician's mind as SUSY models have many resemblances with realistic models but their behaviors in the quantum regime is more under control than the non SUSY case\footnote{This niceness can go up to the case of not having quantum corrections at all, leaving aside possible instantonic contributions.}. A case in point is the celebrated Seiberg-Witten (SW) model proposed in 1994  \cite{swqcd} of a four dimensional SUSY gauge theory that is a fully integrable system, a tremendous breakthrough in mathematical physics.

My PhD thesis \cite{thesis2} was dedicated to the study of the Noether symmetries of this model, the reason being that the SUSY central change $Z$ here plays a role of paramount importance. In a nutshell, (i) $Z$ allows for SSB of the gauge symmetry within the SUSY theory, (ii) it gives the complete and exact mass spectrum (i.e. it fixes the masses for the elementary particles as well as the collective topological excitations), (iii) it exhibits an explicit SL(2, $\Z$) duality symmetry whereas this symmetry is not a symmetry of the theory in the Noether sense, (iv) in the quantum theory it is the most important global piece of information, therefore it is vital for the exact solution of the model. Although Seiberg and Witten had a generic argument on how this quantity should look like in the quantum phase \cite{swqcd}, the first explicit computation was given in {\bf \cite{swplb}} and is reported here. Such computation was very difficult to perform due to a series of complications with the SUSY Noether currents for such a model and in general that we had first to solve (see note \ref{susynote}), the most important being to find the proper improvement terms \cite{thesis2}, {\bf \cite{swannals}} (see \cite{lorsw} for a review).

The recipe I found for computing Noether currents for theories with a non-standard phase space (as it is the case for effective theories like the SW's, in particular in presence of Lagrange multipliers called dummy fields) proved itself to be very useful also in different contexts such as
gauge theories on noncommutative spaces, as we showed in {\bf \cite{NCNoether}} and {\bf \cite{NCNother2}}. Before introducing this new direction in the next Subsection let me conclude this part by deepening a bit more the comments made earlier on the central charge $Z$ (for more details see \cite{thesis2} and \cite{lectnotes1}).

The central charge of the SW model is different from zero and equal to
\begin{equation}\label{zizi}
Z = i\sqrt2 (n_e a + n_m a_D) \;,
\end{equation}
only when $a$, the vacuum expectation value (vev) of the Higgs field $\phi$, and $a_D$, the vev of the {\it dual} of the Higgs field $\phi_D$ are nonzero {\bf \cite{swplb}} (here $n_e$ and $n_m$ are integers denoting the electric and magnetic charges, respectively). What ``dual'' means let me say it later. It is clear then that it is only when the spontaneous symmetry breaking (SSB) of the  gauge symmetry, SU(2) for the original SW model, has taken place that the central charge becomes an important quantity\footnote{On SSB in the classical as well as quantum regime and the related spaces of gauge inequivalent vacua (moduli spaces) see the detailed discussion in my PhD thesis \cite{thesis2}}. In this case, for those SUSY states that satisfy $M = |Z|$, or Bogomolnyi-Prasad-Sommerfield (BPS) states, the mass spectrum is given by
\begin{equation}\label{mass}
M = \sqrt2 |n_e a + n_m a_D| \;,
\end{equation}
and this condition is crucial in order for the number of fields' degrees of freedom to match before and after the SSB: if this does not happen then the whole story stops at massless fields with a trivial vacuum manifold. The formula (\ref{mass}) is called the BPS mass formula and holds for the whole spectrum, elementary excitations (two $W$-like bosons and two fermions in the SU(2) model) as well as topological excitations, monopoles and dyons (the latter being topological excitations carrying electric and magnetic charge at once). For instance the mass of the $W$-like bosons and of the two fermions can be obtained by setting $n_e = \pm 1$ and $n_m = 0$, which gives\footnote{Of course the masses of real $W$ bosons and real quarks are not equal and for this model to make contact with real physics a mechanism to break SUSY needs to be implemented and little is known on how to do that.} $m_W = m_{\rm fermi} = \sqrt2 |a|$, whereas the mass of a monopole ($n_e =  0$ and $n_m = \pm 1$) amounts to $m_{\rm mon.} = \sqrt2 |a_D|$. Incidentally, BPS states play an important role in the search for the SUSY completion of the gravitational CS term \cite{djt} we have found in {\bf \cite{sucs}}. In that work we exploited the topological nature of BPS states in order to understand how to fit the topologically non-trivial (kink) solution of the gravitational CS term we had found in {\bf \cite{cs3}}.

I can now briefly introduce the key ingredient for the exact solution of the SW model and explain some of the objects introduced earlier without a definition: electro-magnetic (em) duality. Let me write $Z$ as
\begin{equation}
Z = i \sqrt2 (n_m , n_e) \left( \begin{array}{c} a_D\\ a \end{array} \right)
\end{equation}
If we act with $S^{-1} \equiv \left( \begin{array}{cc}
0 & - 1
\\ 1 & 0
\end{array} \right)$ on the row vector $(n_m , n_e)$, we exchange electric charge with magnetic charge and {\it vice-versa}. This is
the e.m. duality transformation: it maps electrically charged elementary particles to magnetically charged collective
excitations. Under such transformation all the BPS states are the same. The mass of all the particles has to be given by the
mass formula (\ref{mass}). Therefore to $S^{-1}$ acting on $(n_m , n_e)$ it has to correspond $S$ acting on the column vector, namely
\begin{equation} \label{duality}
\left(
\begin{array}{c} a_D\\ a \end{array} \right) \to S \left(
\begin{array}{c} a_D\\ a \end{array} \right) = \left(
\begin{array}{c} a\\ -a_D \end{array} \right)
\end{equation} so that $Z$ is left invariant.  The mass formula is actually invariant under the
full group $SL(2, \Z)$ of $2 \times 2$ unimodular matrices with integer entries, generated by \footnote{This group structure arises due to two
non-Noether symmetries of the low-energy effective action: the $S$-symmetry in (\ref{duality}), as seen acting on the chiral super-field and its dual, and the $T$-symmetry, again acting on the same fields but this time changing the instanton number $\nu$ as $\nu \to \nu b$, hence forcing $b \in \Z$. On this see \cite{swqcd}, \cite{bilal} and my PhD thesis \cite{thesis2}.}
\be S = \left( \begin{array}{cc} 0 &
1  \\ - 1 & 0 \end{array} \right) \quad {\rm and} \quad T = \left(
\begin{array}{cc} 1 & b
\\ 0 & 1 \end{array} \right) \ee where $b \in \Z$.

The $S$ duality transformation (\ref{duality}) explains why we called $a_D$ the dual of $a$. What it does not properly clarifies is that,
in general (i.e. already for the monopoles Dirac posited for Maxwell electrodynamics) these kind of transformations map a theory written in terms of fields into a theory written in terms of dual of those fields and, since the gauge couplings in the duality-connected regimes are $q_e \sim 1/q_m$ (see, e.g., \cite{lectnotes3}), strong-coupling can be mapped to weak-coupling, solved and mapped back to strong-coupling. Essentially that is why the SW model is solvable\footnote{Of course, the solvability of the model relays with the fact that the low-energy effective lagrangian density is a holomorphic function. Nonetheless, that fact alone does cannot lead to the exact solution (see, e.g., \cite{seiberg}). It is the duality
we just described that plays the key role for reducing the problem to a solvable Riemann-Hilbert problem. For details see my PhD thesis \cite{thesis2}.}.

\subsection{\sc Noncommutativity}

Suppose that, due to a fundamental length $\sqrt{\theta}$, spatiotemporal coordinates do not commute. One way of making this explicit is
\begin{equation}  \label{1}
x^\mu * x^\nu - x^\nu * x^\mu = i \theta^{\mu \nu} \;,
\end{equation}
where $\theta^{\mu \nu}$ is $c$-number valued and constant and the (Moyal-Weyl $*$-)product of any two fields $\phi(x)$ and $\chi(x)$ is defined as $(\phi * \chi) (x) \equiv e^{ \frac{i}{2} \theta^{\mu \nu} \partial^x_\mu \partial^y_\nu} \phi (x) \chi (y)|_{y \to x}$, and it is a suitable generalization of the multiplication law in the presence of a nonzero $\sqrt{\theta}$. The expression (\ref{1}) is not the most general way noncommutativity could take place. For instance, two equally valid, if not more general, approaches are the Lie-algebraic and the coordinate-dependent ($q$-deformed) formulations \cite{wess}, and many other approaches exist. Nonetheless, the canonical form is surely the most simple and the basic features of noncommutativity are captured in this model \cite{lectnotes2}.

The most convincing conjecture for the existence of such a fundamental length and of why this should lead to noncommuting coordinates was given in \cite{doplicher1}. It is based on the generalization to GR of Landau's argument for the special relativistic modifications of the uncertainty Heisenberg relations \cite{landau1}. In a nutshell, the idea of \cite{doplicher1} is that if $\Delta x$ is the uncertainty in the localization of a given event, then $\Delta x$ {\it alone} cannot be smaller than $\sqrt{\theta}$ that is the Schwarzschild radius (assuming spherical symmetry) associated with the energy necessary for the localization of the event because beyond that there would be no visible signal. Since we assumed spherical symmetry $\Delta x = \Delta y = \Delta z$ hence $\Delta x \Delta y \ge \theta$, etc. It is easy to see that with this argument
$\sqrt{\theta} =  \sqrt{G \hbar/ c^3} = \ell_{\rm Planck}$.

Noncommuting coordinates are more than a conjecture (convincing or not) for certain physical systems whose phase-space is truncated in certain ways. In those cases noncommutativity is a testable fact of nature, although not a fundamental one. Essentially, the physical systems I am referring to are constrained systems for which a coordinate (say $y$) becomes the conjugate momentum of another coordinate (say $x$), hence already at the classical level the Poisson (or Dirac) brackets are unusual, $\{ x, y \} = 1$. This sometimes happens in planar systems (2+1 dimensions) and the interaction term can be seen as a limit of the CS term of gauge theory. For an introduction to these physical instances of noncommuting coordinates see Jackiw's lecture \cite{jackiwncft}.

As said in the Prologue, the proposal for noncommuting coordinates has an impact on various aspects of Poincar\'e symmetry. Baring the last paragraph in mind it is then not surprising that the paper that probably pioneered the field now known as ``Lorentz violation'' is  that of 1990 by Carroll, Field and Jackiw \cite{jackiw}, where the authors introduced a Lorentz violating term of CS origin into Maxwell electrodynamics and extensively studied the possibility of visible signatures.

Another link between noncommutativity and Lorentz violation was provided in 1999 by Seiberg and Witten in \cite{seibergwitten}. There they discovered a particular low-energy limit of ST that leads to certain NCFTs that keep standard gauge invariance. This was done by providing a map that gives the noncommutative nonlocal gauge field (transforming under the noncommutative gauge transformations) in terms of the standard commutative local gauge field (transforming under the usual gauge transformations) and of the noncommutative parameters $\theta_{\mu \nu}$. This map is known as the Seiberg-Witten (SW) map\footnote{ST is not necessary to construct gauge theories on noncommutative spaces and not even to obtain the SW map. This is shown in \cite{wess} where $[x , \cdot ~]$ is taken as a derivative and made it gauge covariant.}. That these theories are Lorentz symmetry violating has been debated for some time but was clearly shown in my work {\bf \cite{NCNoether}}, {\bf \cite{NCNother2}} that indeed they are.

Two are the directions I have pursued in my research in the area. On the one hand, as mentioned, I focused on the clear understanding of the spatiotemporal symmetry properties of these theories {\bf \cite{NCNoether}}, {\bf \cite{NCNother2}}, along the lines of the application of the Noether procedure developed for SUSY theories with dummy field \cite{thesis2}, {\bf \cite{swannals}}. On the other hand I hunted for experimental evidences of fundamental noncommutativity in simple theoretical set-ups {\bf \cite{NCsync}}, \cite{NCsync2}, {\bf \cite{cerenkov}}.

Noncommutativity is probably the second most used theoretical frame to study Lorentz violation\footnote{The quantum phase of NCFTs is still an open issue: it was shown that novel divergencies appear in such theories \cite{minwallaseiberg}, but whether these are unavoidable or not is still to be clarified. Lately it was argued that such divergencies should not be present if noncommutativity is implemented via a deformed (twisted) coproduct \cite{balachandran}}, the first being Colladay and Kostelecky's Standard Model Extension (SME) \cite{colladaykostelecky}. It was later proved that indeed the two approaches can be actually reconciled \cite{Carroll:2001ws} and the key point is that the Lorentz (and possibly CPT) violating terms added to the Standard Model (or gravity) terms have the form
\begin{equation}\label{SME1}
C^{(k)}_{\;\;\;\;\mu ... \nu} \; \times \; ({\rm SM \; fields \; and \;
derivatives})^{\mu ... \nu} \;,
\end{equation}
where $C^{(k)} \sim 1/m_{\rm Planck}^k$, so that the higher one goes with the power $k$, the smaller the effect.

The idea here is to find physical set-ups where the Planck size effects can be magnified to the point of giving rise to signatures measurable by the instruments at our disposal. Those effects should have visible signatures ``over here'', i.e. at reachable energy scales too. This is quite an interesting point of view and is somehow opposed to the idea that it is impossible to test phenomena that take place at the Planck scale.

The theoretical laboratory we used to pursue both directions, the study of spatiotemporal properties in general and the hunt for experimental signatures, has been mainly Noncommutative Electrodynamics (NCED), whose action is
\begin{equation}  \label{ncym} \hat{I} = \int d^4 x \hat{\cal L} = -
\frac{1}{4} \int d^4 x \hat{F}^{\mu \nu} \hat{F}_{\mu \nu} \;,
\end{equation}
where $\hat{F}_{\mu \nu} = \partial_\mu \hat{A}_\nu - \partial_\nu \hat{A}_\mu - i [ \hat{A}_\mu , \hat{A}_\nu ]_* $. As said, the
nonlocal field $\hat{A}_\mu$ can be expressed in terms of a standard U(1) gauge field $A_\mu$ and of $\theta^{\mu \nu}$ by
means of the SW map\footnote{$\hat{A}_\mu (A,\theta) \to A_\mu$ as $\theta^{\mu \nu} \to 0$, hence, in that limit, $\hat{F}_{\mu \nu}
\to F_{\mu \nu} = \partial_\mu A_\nu - \partial_\nu A_\mu$.} \cite{seibergwitten} $\hat{A}_\mu (A,\theta)$, that at $O(\theta)$
reads
\begin{equation}
\hat{A}_{\mu}(A, \theta) = A_{\mu} - \frac{1}{2} \theta^{\alpha
\beta}A_{\alpha}(\partial_{\beta}A_{\mu} + F_{\beta \mu})  \;.
\end{equation}

The Noether currents for space-time transformations of the theory (\ref{ncym}) are $J_f^\mu = \Pi^{\mu \nu} \delta_f A_\nu -
\hat{\cal L} f^\mu$, where $\Pi^{\mu \nu} = \delta \hat{\cal L} / \delta \partial_\mu A_\nu$, and for translations (the only symmetric
case) $f^\mu \equiv a^\mu \equiv {\rm const}$. Hence the (conserved) energy-momentum tensor $T^{\mu \nu} = \Pi^{\mu \rho}
F^\nu_\rho - \eta^{\mu \nu} \hat{\cal L}$ is in general not symmetric, a clear sign of the breaking of Lorentz symmetry
{\bf \cite{NCNoether}}, {\bf \cite{NCNother2}}. This point has been (and still is) the subject of a debate that can be summarized with the question:
should $\theta^{\mu \nu}$ be transformed under spatiotemporal transformations according to its indices structure\footnote{I.e. as a rank two tensor, see the general discussion in the first part of Subsection \ref{confsusy}.} or should it not be transformed at all? Our point of view on this is that
$\delta \theta^{\mu \nu} = 0$, an instance only compatible with translational symmetry and with the subgroup $SO(2) \times SO(1,1)$ of the Lorentz group, see {\bf \cite{NCNoether}} and especially {\bf \cite{NCNother2}}. Indeed the analogy we were able to make explicit between $\theta^{\mu \nu}$ and the dummy fields of SUSY was pivotal in clarifying this issue.

Conservation of $T^{\mu \nu}$ means conservation of the Poynting vector
\begin{equation}\label{poynting}
  \vec{S} = \frac{c}{4 \pi} \vec{D} \times \vec{B} =
  \frac{c}{4 \pi} \vec{E} \times \vec{H} \;,
\end{equation}
where $D^i \equiv \Pi^{i 0}$ and $H^i \equiv \frac{1}{2} \eps^{i j k} \Pi_{j k}$ are the constitutive relations containing all the relevant information about the Lorentz violating vacuum.

The effects of a nonzero $\theta$, if any, are very small, thus the $O(\theta)$ model would do the job
\begin{equation}  \label{othetamaxwell}
\hat{I} = - \frac{1}{4} \int d^4 x \; [F^{\mu \nu} F_{\mu \nu}
-\frac{1}{2} \theta^{\alpha \beta} F_{\alpha \beta} F^{\mu \nu}
F_{\mu \nu} + 2 \theta^{\alpha \beta} F_{\alpha \mu} F_{\beta \nu}
F^{\mu \nu}] + J_\mu \hat{A}^\mu \;,
\end{equation}
where the vector field is coupled to an external current, and the $O(\theta)$ SW map and $*$-product are used. In {\bf \cite{NCsync}} we studied the
synchrotron radiation in NCED, and indeed found the amplification we were looking for. It is matter of solving the modified
Maxwell equations descending from (\ref{othetamaxwell}) with the settings: I. charged particle moving (circularly) in the plane
$(1,2)$: $J_\mu = e c \beta_\mu \delta(x_3) \delta^{(2)} (\vec{x} - \vec{r}(t))$, where $\vec{r}(t)$ is the position of the
particle, and $\beta_\mu = (1, \vec{v}/c)$; II. the background magnetic field along the $x_3$-direction, $\vec{b} = (0, 0, b)$; III. similarly for the spatial part of $\theta_{\mu \nu}$, i.e. $\vec{\theta} = (0, 0, \theta)$, with $\theta^i = \frac{1}{2} \epsilon^{i j k} \theta_{j k}$.

Keeping only contributions of order $O(e/R)$, where $R$ is the distance from the source, one can compute the electric and magnetic fields (that
have quite involved expressions) and from them compute the spectrum in the ultra-relativistic regime and far from the source. The ratio of the (Fourier transform of the) $\theta$-corrected radiated energy $d I (\omega) / d \Omega$ to the $\theta = 0$ one is found in {\bf \cite{NCsync}} to be
\begin{equation}\label{ratio}
X \equiv \frac{d I (\omega) / d \Omega}{d I(\omega) / d
\Omega|_{\theta = 0}} \sim 1 + 20 \left( \frac{\omega_0}{\omega} \right)^{\frac{2}{3}}
\theta \; b \;\gamma^4 \;,
\end{equation}
for $\omega_0 << \omega << \omega_c$, where $\omega_0$ and $\omega_c$ are the two characteristic frequencies in the ultra-relativistic approximations for the synchrotron, the cyclotron ($\omega_0 \sim c / |\vec{r}|$) and the critical ($\omega_c = 3 \omega_0 \gamma^3$) frequencies, respectively. Thus (\ref{ratio}) provides an example of how the effects of noncommutativity, $\lambda = 2 \theta b$, can be amplified in certain physical set-ups, in this case by the fourth power of $\gamma$.

Unfortunately, the limits on the velocity of earth-based electron-synchro-trons, hence in turn of the associated magnetic field $b$, constraint the effects to $X \sim 1 + 10^{-10}$. Furthermore, within NCED, at present these numbers cannot be ameliorated even by including synchrotron radiation of astrophysical origin and vacuum \v{C}erenkov radiation {\bf \cite{cerenkov}} (see also \cite{NCsync2}). Other directions, though, appear to be more promising as the preliminary results of \cite{colorhorizon} show.

Other investigators have also looked into the synchrotron radiation within the SME \cite{altschul} and other Lorentz violating models \cite{urrutia}. They find similarly important departures from the Lorentz preserving formula in agreement with the above presented modification of the energy spectrum and with a previously proposed formula for the maximum frequency based on kinematical general arguments \cite{jacklibmat}.

Synchrotron radiation is just one among a plethora of phenomenological setups that have been investigated, theoretically and experimentally, in the last two decades of search for Lorentz violation. Despite that no signals of violation have been found. In my view, though, this search, that keeps theoreticians and experimentalists side by side, is important because it might serve the scope of changing in an unsuspected fashion some of the established paradigms of the theoretical investigation, one of these being SUSY. For instance, already Berger and Kostelecky \cite{susykostelecky} have investigated, within the SME, the effects of Lorentz violation on SUSY and have found that a (modified) SUSY algebra can be written. Similarly, the effects of noncommuting coordinates on SUSY has been considered \cite{susyrivelles} and some preliminary results were the outcome of the MSc thesis work of one of my students \cite{vydrova}. An interesting unanswered question in this respect is: What if we do not assume SUSY but reconsider the no-go theorems with the assumption that the spatiotemporal group is no longer ISO(3,1)?

\section{\sc Gravity at the intersection}

From the previous Sections it should be easy to see why the two research activities just described both have natural applications in gravity and gravity-like models. This was explicitly said in various occasions. Our findings in the exploration of the thermal/statistical properties of the QFT vacuum and their description in terms of q-groups (see Section~2) opened-up the way to a novel systematic approach to quantum thermal phenomena induced by the presence of an horizon {\bf \cite{qBH1}}, to the introduction in this context of an (holographic) entropy operator {\bf \cite{TFDBH}} and later to the proposal of a gravity analog of thermal hadron production in high energy scattering processes {\bf \cite{rindler}}. Our expertise with conformal symmetry in flat and curved space-times and with SUSY (see Subsection~3.1) was pivotal for the discovery of a kink solution of the CS (or conformal) gravitational term {\bf \cite{cs3}} and for the investigation of the SUSY completion of the CS gravitational term and for the study of the BPS nature of the SUSY version of that kink {\bf \cite{sucs}}. As for the development of these directions, most of my work in progress is dedicated to pursue farther these findings along the lines of gravity analogs in elastic defects systems \cite{gaugevirus}, in two-dimensional rippled graphene \cite{weylgraphene}.

Since I believe that in Section 2 I provided enough introduction to the work in {\bf \cite{qBH1}}, {\bf \cite{TFDBH}}, what I would like to do here is to briefly introduce the analogies and differences between standard gauge theories and gravity theories\footnote{The literature on the subject is vast and with various points of view. A succinct and insightful account is in \cite{lor1979}. See also \cite{ryder}. }. Some of this is already presented in {\bf \cite{cs3}}, here I should try to put a bit more flash on those bones. This should serve the double scope of introducing the language of the work in {\bf \cite{cs3}} and {\bf \cite{sucs}}, and of explaining how concepts born in a gravitational context can be used in other areas of physics, hence to point to some of my work in progress. Most of what I shall say is part of the lectures I regularly deliver \cite{lectnotes3}.

In GR the covariant derivative\footnote{I omit the ``group indices'' in (\ref{dgamma}) just like in (\ref{dA}).}
\begin{equation}\label{dgamma}
    \nabla_\mu = \partial_\mu + \Gamma_\mu \;,
\end{equation}
resembles in any respect the covariant derivative of gauge theories
\begin{equation}\label{dA}
    D_\mu = \partial_\mu + A_\mu \;.
\end{equation}
Indeed in both cases $\Gamma_\mu$ and $A_\mu$ are the response to a parallel transport in space-time, for (\ref{dgamma}) acting on the space-time itself, for (\ref{dA}) acting on the Lie group G. Let us then formally write
\begin{equation}\label{gammaA}
\Gamma^\lambda_{\;\; \mu \nu} \equiv \left( A_\mu \right)^\lambda_{\; \; \nu} \;.
\end{equation}
The analogy continues when we look at how this gauge field transforms under general coordinate transformations
\begin{equation}\label{Atrnsf}
\left( A'_\mu \right)^\lambda_{\; \; \nu} (x') = \frac{\partial x^\rho}{\partial {x'}^\mu}
\left[ U^{-1} A_\rho (x) U + U^{-1} \partial_\rho U \right]^\lambda_{\; \; \nu} \;,
\end{equation}
where
\begin{equation}\label{UUinv}
    U^\tau_{\; \; \nu} = \frac{\partial x^\tau}{\partial {x'}^\nu} \quad {\rm and} \quad
    (U^{-1})^\lambda_{\; \; \rho} = \frac{\partial {x'}^\lambda}{\partial x^\rho} \;,
\end{equation}
are the gauge functions. Thus (\ref{Atrnsf}) is a general coordinate transformation acting on the index $\mu$ of the
``gauge transformed'' ${A'}_\mu = U^{-1} A_\mu (x) U + U^{-1} \partial_\mu U$, that is the standard gauge transformed with respect to the other indices $\lambda, \nu$. Of course some open-mindedness is necessary here as $U$ itself is a general coordinate transformation.

We can go farther with the analogy by considering the curvature tensor and realize that it is the field strength of $A_\mu$ in (\ref{gammaA})
\begin{eqnarray}\label{RasF}
R^\lambda_{\;\; \sigma \mu \nu} & = & \partial_{[\mu} \Gamma^\lambda_{\;\; \nu] \sigma}
+ \Gamma^\lambda_{\;\; \mu \rho} \Gamma^\rho_{\;\; \nu \sigma} - \Gamma^\lambda_{\;\; \nu \rho} \Gamma^\rho_{\;\; \mu \sigma} \\
& \equiv & \left( \partial_{[ \mu} A_{\nu]} + [A_\mu , A_\nu ] \right)^\lambda_{\; \; \sigma} =
\left( F_{ \mu \nu} \right)^\lambda_{\; \; \sigma} \;.
\end{eqnarray}
Here, though, the analogy stops because (i) $A_\mu$ in gauge theories is the primary object, while in standard GR, i.e. when metricity is imposed, $\Gamma^\lambda_{\;\; \mu \nu}$ is a derived object, the primary one being the metric $g_{\mu \nu}$; (ii) the Lagrangian for gauge theories is (proportional to) the trace of the square of the field strength
\begin{equation}\label{Lgauge}
    {\cal L}_{\rm gauge} \sim {\rm Tr} (F_{\mu \nu} F^{\mu \nu}) \;,
\end{equation}
while in GR it is the trace of the field strength
\begin{equation}\label{LGR}
    {\cal L}_{\rm GR} \sim \left( F_{ \mu \nu} \right)^{\mu \nu} \;;
\end{equation}
(iii) the gauge group $G$ for gauge theories is not fixed by the space-time, while for GR is $G = SO(n-1, 1)$ in $n$ dimensions and the translations need to be identified with the diffeomorphisms; (iv) while in gauge theories the force is proportional to $F_{\mu \nu}$, in GR, due to the equivalence principle, it is proportional to the potential $\Gamma^\lambda_{\;\; \mu \nu}$.

The first mismatch in the list can be solved by the formulation of gravity in terms of vielbeins $e^a_\mu$, i.e. objects that
carry an Einstein index $\mu$ (responding to general coordinate transformations) along with a Lorentz index $a$ (that only responds
to local Lorentz transformations), defined by\footnote{One way of understanding $e^a_\mu$ is to see them as a frame at every point of the curved manifold where general covariant quantities
become flat, e.g.
\begin{equation}\label{vflat}
    V^\mu e^a_\mu = V^a \;,
\end{equation}
with $V^a$ a vector under Lorentz transformations and a set of four scalars under general coordinate transformations. The inverse of the
vielbein, $E_a^\mu$, can also be introduced
\begin{equation}\label{inverseE}
    e_\mu^a \, E^\mu_b = \delta^a_b \quad {\rm and} \quad e_\mu^a \, E^\nu_a = \delta^\nu_\mu \;.
\end{equation}
There are important subtleties related to the existence of the inverse vielbein $E_a^\mu$ in the quantum theory as noticed for the first time in \cite{witten}. See also \cite{carlip}.}
\begin{equation}\label{vielbein}
    g_{\mu \nu} = \eta_{a b} e^a_\mu e^b_\nu \;.
\end{equation}
The full covariant derivative of the Vielbein is an important object that introduces the connection ${\omega_\mu} ^a_{\;\; b}$ that takes into account the parallel transport acting on the flat index
\begin{equation}
\nabla_\mu e^a_{\;\; \nu} = \partial_\mu e^a_{\;\; \nu} -
\Gamma^\lambda_{\mu \nu} e^a_{\;\; \lambda} +  {\omega_\mu} ^a_{\;\; b} e^b_{\;\; \nu} \;.
\end{equation}
Thus, when an appropriate condition is satisfied, the pair of variables $(g_{\mu \nu} , \Gamma^\lambda_{\;\; \mu \nu})$ can be replaced by $(e^a_\mu , {\omega_\mu} ^a_{\;\; b})$. The condition is
\begin{equation}\label{cartan}
   \nabla_\mu e^a_\nu = 0 \;,
\end{equation}
and it eliminates $\Gamma$ in favor of $\omega$, $\Gamma^\lambda_{\;\; \mu \nu} = E_{\;\; a}^\lambda  {\omega_\mu} ^a_{\;\; b} e^b_{\;\; \nu} + E_{\;\; a}^\lambda \partial_\mu e^a_{\;\; \nu}$, hence, through (\ref{RasF}), it gives
\begin{equation}\label{fisrtCartan2}
(R_{\mu \nu})^a_{\; \; b} = \left( \partial_{[\mu} \; \omega_{\nu]} + \omega_{[\mu} \; \omega_{\nu]} \right)^a_{\;\; b} \;,
\end{equation}
where $R^\rho_{\; \sigma \mu \nu} = E^\rho_{\;\; a} (R_{\mu \nu})^a_{\; \; b} e^b_{\;\; \sigma}$. By defining the torsion as $T^\lambda_{\;\; \mu \nu} = \Gamma^\lambda_{\;\; [ \mu  \nu]} = E_{\;\; a}^\lambda T^a_{\;\; \mu \nu}$ the condition (\ref{cartan}) gives
\begin{equation}\label{secondCartan2}
T^a_{\;\; \mu \nu} = \left( \partial_{[\mu} \; e_{\nu]} + \omega_{[\mu} \; e_{\nu]} \right)^a \;.
\end{equation}

Equations (\ref{fisrtCartan2}) and (\ref{secondCartan2}) are called \textit{Cartan structure equations} and have a deep geometrical meaning that reaches far beyond GR as can be immediately recognized by the presence of torsion. It is only when torsion is zero that standard GR is recovered and the sole variable left is $e^a_\mu$ as $T = 0 \to \omega (e)$. To see how this formalism makes the first step in the gauge/gravity theory identification one needs to consider the following expression as the gauge field for gravity, instead of (\ref{gammaA})
\begin{equation}\label{gaugeAgravity}
    A_\mu = e^a_\mu P_a + {\omega_\mu} ^{a  b} M_{a b} \;,
\end{equation}
where $P_a, M_{a b}$ are the generators of the Poincar\'e group. Hence, the program consists in identifying $e^a_\mu$ as the gauge field associated to translations (diffeomorphisms) and ${\omega_\mu} ^{a  b}$ as the gauge field associated to local Lorentz. An important point is to keep torsion, hence to have $(e^a_\mu , {\omega_\mu} ^a_{\;\; b})$ as independent variables {\it off shell}, i.e. before the implementation of the Euler-Lagrange (EL) equations descending from the EH action written as
\begin{equation}\label{nearlyCS}
I_{\rm EH}= \int d^4 x \sqrt{|g|} R =
\frac{1}{4} \int d^4 x \eps^{\mu \nu \lambda \kappa} \eps_{a b c d} \; e^a_\mu \; e^b_\nu \; (R_{\lambda \kappa})^{c d} \;,
\end{equation}
that indeed gives the vacuum Einstein equations when varied with respect to $e$ and $T = 0$ when varied with respect to $\omega$. Thus (\ref{gaugeAgravity}) and (\ref{nearlyCS}) give as primary objects of gravity the potentials, just like in gauge theories.

This formulation, called the {\it first order formulation}\footnote{This name comes from the fact that the EL equations are first order differential equations.}, is also general enough to include cases outside standard GR that keep GR in a limit. For instance, when SUSY is invoked within a gravitational context (SUGRA) the first order formalism reveals itself to be the most convenient \cite{wb} and that is precisely what we have used in {\bf \cite{sucs}}. In that case a SUSY partner of the vielbein (a Rarita-Schwinger spin-3/2 field $\psi^\alpha_\mu$) needs to be introduced, $\delta_{\rm SUSY} {e^a_\mu} \sim \eps \gamma^a \psi_\mu$, and the spin connection, even when it is required to be a {\it dependent} variable, besides the contribution coming from the $e$-dependence, acquires a fermionic contribution, $\omega(e, \psi)$, that always gives a nonzero torsion \cite{wb}.

Another application outside canonical gravity is that to topological defects in elastic media that can be performed in 2+1 dimensions \cite{katanaev}, \cite{kleinert}. In a nutshell, it is possible to describe {\it dislocation} defects by keeping torsion but by requiring zero curvature, $T \neq 0$ and $R = 0$, and to describe {\it disclination} defects by doing the opposite, $T = 0$ and $R \neq 0$. Mixed situations are possible as well. The Burgers vectors (that describe dislocations) are $e^a_\mu$, the Frank vectors (that describe disclinations) are $\omega^a_\mu$. We are pursuing that direction in our work in progress \cite{gaugevirus}.

The aware reader might have noticed that the connection lost one index. This is one of the peculiarities of the three dimensions
\begin{equation}\label{3dpeculiarity}
 {\omega_\mu}^{a  b} = \eps^{a b c} \omega_{c \; \mu} \;,
\end{equation}
that, in subtle ways, as shown by Witten in \cite{witten}, has far reaching consequences one being the solution of the second and third mismatches between gravity and gauge theories that only works if the space-time is three dimensional. Recall that the mismatches I am referring to are the form of the action and the nature of the transformations (diffeomorphisms and local Lorentz on the gravity side, gauge transformations $\delta A_\mu = D_\mu u$ on the other side, where $u = \rho^a P_a + \tau^a M_a$). Indeed in \cite{witten} it is proved that GR in three dimensions is a CS gauge theory with action
\begin{equation}\label{EHCS}
I_{\rm EH}= \int d^3 x \; \eps^{\mu \nu \lambda} \; \left(\frac{1}{2} e^a_\mu \; \partial_\nu \; \omega_{a \lambda}
+  \frac{1}{3} \eps_{a b c} \; e^a_\mu \omega^b_\nu \omega^c_\lambda \right) \;,
\end{equation}
with gauge group $ISO(2,1)$.

In previous investigations Deser, Jackiw and Templeton \cite{djt} showed that in three dimensions another term has to be considered besides
the standard EH term, and it happens that this term as well has a CS structure\footnote{There are many subtleties that we are not mentioning here as they are well explained in literature, see \cite{djt} but also our own work {\bf \cite{cs3}}. For instance, there is no need to use the first order formulation to see the CS term in three dimensions, as opposed to what happens for the EH part where only the first order formulation shows the CS structure. Indeed a term of the form
\begin{equation}
I_{\rm CS} (\Gamma) =  \int d^3 x \eps^{\mu \nu \lambda} \left( \frac{1}{2} \Gamma^\rho_{\mu \sigma} \partial_\nu
\Gamma^\sigma_{\lambda \rho} + \frac{1}{3} \Gamma^\rho_{\mu
\sigma} \Gamma^\sigma_{\nu \tau} \Gamma^\tau_{\lambda \rho} \right) \;,
\end{equation}
exists within the second order formulation, the difference with the $I_{\rm CS}$ in (\ref{CSCS}) being the winding of the Dreibein that, although it leads not to a quantization like in certain standard gauge theories and although it does not affect the EL equations, it is necessary to have invariance under local Lorentz.}
\begin{equation}\label{CSCS}
I_{\rm CS} = \int d^3 x \eps^{\mu \nu \lambda}
\left( \frac{1}{2} \omega^a_{\mu} \partial_\nu \omega_{a \lambda} + \frac{1}{3}
\eps_{a b c} \omega^a_\mu \omega^b_\nu \omega^c_\lambda \right) \;.
\end{equation}
The full theory in three dimensions is then (leaving aside contributions from a cosmological constant)
\begin{equation}
I_{\rm EH} + \frac{1}{M} \; I_{\rm CS}
\end{equation}
that leads to a massive $(M)$ scalar graviton and it is called Topologically Massive Gravity (TMG) \cite{djt}.

Since its discovery TMG is a very active field of research as can be seen from the contributions to a work-shop held in Vienna last year \cite{viennatmg}. Our own contribution to this field has been the discovery of a kink solution {\bf \cite{cs3}} of the $M \to 0$ limit of TMG, i.e. of that sector of TMG governed by the CS gravity term alone, and later the systematic study of the SUSY of the kink and of all other solutions {\bf \cite{sucs}}. The work in {\bf \cite{cs3}} immediately ignited further research and contributed significantly to the revived interest in the field of TMG.

A few last remarks are in order because we often referred to the theory $I_{\rm CS}$ as conformal gravity and from the above it is not clear why this name makes sense. First the EL equations descending  from (\ref{CSCS}) are
\begin{equation}\label{cotton}
C^{\mu \nu} = - \frac{1}{2 \sqrt{|g|}} \left( \eps^{\mu \lambda \kappa} \nabla_\lambda R_\kappa^\nu
+ \eps^{\nu \lambda \kappa} \nabla_\lambda R_\kappa^\mu  \right) = 0 \;,
\end{equation}
where $C^{\mu \nu}$ is the Cotton tensor and, being traceless, indicates conformal invariance of (\ref{CSCS}). Second, the solutions to the EL equations (\ref{cotton}) are all conformally flat, $g_{\mu \nu} = \sigma \eta_{\mu \nu}$, as in three dimensions the Cotton tensor plays the role of the Weyl tensor in four dimensions\footnote{The Weyl tensor is identically zero in three dimensions. This is another peculiarity of the three dimensions and it leads to the absence of propagating degrees of freedom \cite{djth} unless the full TMG is considered \cite{djt}.}. Third, it was proved in \cite{hornewitten} that, in the gauge where the Vielbein is everywhere invertible, the theory (\ref{CSCS}) is a proper gauge theory of the group $SO(3,2)$, that is the conformal group in three dimensions. The gauge field in this case is a straightforward generalization of (\ref{gaugeAgravity})
\begin{equation}\label{gaugeAgravityconf}
    A_\mu = e^a_\mu P_a + \omega_\mu^a M_a + \lambda_\mu^a K_a + \phi_\mu D \;,
\end{equation}
where, besides the generators of the Poincar\'e group, there are $K_a$ and $D$, generators of the special conformal transformations and of dilations, respectively.

\section*{\sc Acknowledgments}

I am indebted to all my coworkers. I learned from each of them. Among the others, I learned from Peppino Vitiello that the guiding stars to do this job are joy and intercontinental movings, from the late Lochlainn O'Raifeartaigh how to properly spell that joy and his name, from Roman Jackiw how Boston winters cannot stop a physicist to take that joy to the sky and from Siddhartha Sen that to bridge the distance between mathematical beauty and experimental reality is great fun.

Finally, I want to thank Ji\v{r}\'i Ho\v{r}ej\v{s}\'i for the trust he accorded to my intellectual enterprize.

\newpage

\thispagestyle{empty}

\section{\sc Selected Papers}\label{selpap}

In this last Section I collect the 17 most relevant papers grouped into the three areas of research presented earlier. They could be read immediately after the relevant Section above to make the most of the introduction (e.g., the papers in Subsection \ref{selpap}.1 could be read after Section 2 above) but some papers, inevitably, belong to more than one Section.

\subsection{\sc QFT Vacuum and Quantum Groups}

\noindent {\it The vacuum of QFT

\noindent is not empty.

\noindent The interaction picture

\noindent does not exist.

\noindent Quantum groups

\noindent are neither quantum nor groups.

\noindent Can we make some positive statements here?}

\vfill

\noindent (Ref.\cite{qVN}) A.~Iorio, G.~Vitiello, {\it Quantum Groups and von Neumann Theorem},
Mod. Phys. Lett. B {\bf 8} (1994) 269.

\vskip .3cm

\noindent (Ref.\cite{qQD}) A.~Iorio, G.~Vitiello, {\it Quantum Dissipation and Quantum Groups},
Ann. Phys. {\bf 241} (1995) 496.

\vskip .3cm

\noindent (Ref.\cite{neutrino}) E.~Alfinito, M.~Blasone, A.~Iorio, G.~Vitiello, {\it Squeezed Neutrino Oscillations in Quantum Field Theory},
Phys. Lett. B {\bf 362} (1995) 91.

\vskip .3cm

\noindent (Ref.\cite{pla}) E.~Celeghini, S.~De Martino, S.~De Siena, A.~Iorio, M.~Rasetti, G.~Vitiello,
{\it Thermo Field Dynamics and Quantum Algebras}, Phys. Lett. A {\bf 244} (1998) 455.

\vskip .3cm

\noindent (Ref.\cite{shor}) P.~Giorda, A.~Iorio, S.~Sen, S.~Sen, {\it Semi-classical Shor's Algorithm}, Phys. Rev. A {\bf 70} (2004) 032303.

\newpage

\thispagestyle{empty}

\subsection{\sc Spatiotemporal Symmetries: Conformal, SUSY and Noncommutative}

\noindent {\it First stretch your space at pleasure,

\noindent and hope that none has a scale to break your game.

\noindent Then make it super,

\noindent and teach all your fields to be super as well.

\noindent Finally, let depth, height and length be interchangeable only at a price,

\noindent and see if you can measure something!}

\vfill
\noindent (Ref.\cite{weyl}) A.~Iorio, L.~O'Raifeartaigh, I.~Sachs, C.~Wiesendanger, {\it Weyl-Gauging and Conformal Invariance},
Nucl. Phys. B {\bf 495} (1997) 433.

\vskip .3cm

\noindent (Ref.\cite{swplb}) A.~Iorio, {\it A Note on Seiberg-Witten Central Charge}, Phys. Lett. B {\bf 487} (2000) 171.

\vskip .3cm

\noindent (Ref.\cite{swannals}) A.~Iorio, L.~O'Raifeartaigh, S.~Wolf, {\it Supersymmetric Noether Currents and Seiberg-Witten Theory},
Ann. Phys. {\bf 290} (2001) 156.

\vskip .3cm

\noindent (Ref.\cite{NCNoether}) A.~Iorio, T.~Sykora, {\it On the Space-Time Symmetries of Noncommutative Gauge Theories},
Int. J. Mod. Phys. A {\bf 17} (2002) 2369.

\vskip .3cm

\noindent (Ref.\cite{NCsync}) P.~Castorina, A.~Iorio, D.~Zappal\`a, {\it Noncommutative Synchrotron},
Phys. Rev. D {\bf 69} (2004) 065008.

\vskip .3cm

\noindent (Ref.\cite{cerenkov}) P.~Castorina, A.~Iorio, D.~Zappal\`a, {\it On the Vacuum \v{C}erenkov Radiation in Noncommutative
Electrodynamics and the Elusive Effects of Lorentz Violation},
Europhys. Lett. {\bf 71} (2005) 912.

\vskip .3cm

\noindent (Ref.\cite{NCNother2}) A.~Iorio, {\it Comment on ``Noncommutative gauge theories and Lorentz symmetry, Phys. Rev. D {\bf 70} (2004) 125004''},
Phys. Rev. D {\bf 77} (2008) 048701.

\newpage

\thispagestyle{empty}

\subsection{\sc Gravity at the intersection}

\noindent {\it I got here,

\noindent while I was going somewhere else.

\noindent I stay here,

\noindent and I look at something else.

\noindent I see,

\noindent I cannot escape from gravity...}

\vfill

\noindent (Ref.\cite{qBH1}) A.~Iorio, G.~Lambiase, G.~Vitiello, {\it Quantization of Scalar Fields in Curved Background and Quantum Algebras},
Ann. Phys. {\bf 294} (2001) 234.

\vskip .3cm

\noindent (Ref.\cite{cs3}) G.~Guralnik, A.~Iorio, R.~Jackiw, S.-Y.~Pi,
{\it Dimensionally Reduced Gravitational Chern-Simons Term and its Kink},
Ann. Phys. {\bf 308} (2003) 222.

\vskip .3cm

\noindent (Ref.\cite{TFDBH}) A.~Iorio, G.~Lambiase, G.~Vitiello,
{\it Entangled Quantum Fields near the Event Horizon, and Entropy},
Ann. Phys. {\bf 309} (2004) 151.

\vskip .3cm

\noindent (Ref.\cite{sucs}) L.~Bergamin, D.~Grumiller, A.~Iorio, C.~Nu\~{n}ez,
{\it Chemistry of Chern-Simons Supergravity: reduction to a BPS
kink, oxidation to M-theory and thermodynamical aspects}, J. High Energy Phys. {\bf 11} (2004) 021.

\vskip .3cm

\noindent (Ref.\cite{rindler}) D.~Grumiller, A.~Iorio, P.~Castorina,
{\it The Exact String Black Hole behind the Hadronic Rindler Horizon?},
Phys. Rev. D {\bf 77} (2008) 124034.

\end{document}